\documentclass[%
 reprint,
superscriptaddress,
 amsmath,amssymb,
 aps,
]{revtex4-2}

\usepackage{amssymb}
\usepackage{amsmath}
\usepackage{graphicx}
\usepackage{hyperref}
\usepackage{cleveref}
\usepackage{xcolor}

\usepackage{graphicx}
\usepackage{dcolumn}
\usepackage{bm}

\begin{document}

\preprint{APS/123-QED}

\title{Maximum entropy in dynamic complex networks}

\author{Noam Abadi}
 \email{n.abadi@rug.nl}
\author{Franco Ruzzenenti}
 \affiliation{Integrated Research on Energy, Environment and Society, Faculty of Science and Engineering, University of Groningen}

\date{\today}

\begin{abstract}

The field of complex networks studies a wide variety of interacting systems by representing them as networks. To understand their properties and mutual relations, the randomisation of network connections is a commonly used tool. However, information-theoretic randomisation methods with well-established foundations mostly provide a stationary description of these systems, while stochastic randomisation methods that account for their dynamic nature lack such general foundations and require extensive repetition of the stochastic process to measure statistical properties. In this work, we extend the applicability of information-theoretic methods beyond stationary network models. By using the information-theoretic principle of maximum caliber we construct dynamic network ensemble distributions based on constraints representing statistical properties with known values throughout the evolution. We focus on the particular cases of dynamics constrained by the average number of connections of the whole network and each node, comparing each evolution to simulations of stochastic randomisation that obey the same constraints. We find that ensemble distributions estimated from simulations match those calculated with maximum caliber and that the equilibrium distributions to which they converge agree with known results of maximum entropy given the same constraints. Finally, we discuss further the connections to other maximum entropy approaches to network dynamics and conclude by proposing some possible avenues of future research.
\end{abstract}

\maketitle

\section{Introduction}
\label{sec:introduction}

Complex networks is a growing field of research that studies a wide variety of interacting systems, ranging from molecular\,\cite{bertz1981first,GarciaSanchez2022Emergence} to socio-economic scales\,\cite{pedersen2021modeling,merz2023disruption}. Individual components of the system (e.g. atoms, people, or companies) are generally referred to as nodes while interactions between pairs of nodes (e.g. forces, language, or money) are called links. While specific disciplines study these systems in detail, providing empirical and theoretical accounts of the patterns observed in them, many applications of complex networks rely on statistical methods due to their trans-disciplinary nature\,\cite{pastor2003statistical}. One such method is network randomisation, the reconfiguration of which and how different pairs of nodes are  connected. The goal of this is to bring out system characteristics beyond the details of particular case studies\,\cite{newman2003mixing,dadashi2010rewiring, bertotti2020network}. Based on the existing literature on the topic, we distinguish two types of network randomisation: explicit randomisation and distribution sampling.

Explicit randomisation requires considering a stochastic process that modifies the network configuration in steps. A network configuration is a particular arrangement of links among pairs of nodes in a network and the properties of these links (e.g. the magnitude and direction of forces or amounts of money). For example, consider a process that, at every step, randomly chooses a connection in the network and places it between some random disconnected pair of nodes. From any initial network, the randomisation step can be successively applied, defining a trajectory in the space of all possible network configurations. Network configurations obtained from the same trajectory can then be interpreted as states that a time-dependent network takes as it evolves, shedding light on the dynamics of the system. Meanwhile, network configurations from different trajectories after a predefined number of steps (interpreted as the elapsed time) correspond to samples of randomised networks, where the level of randomisation can be tuned by the number of steps. Additionally, the amount of trajectories in which a particular configuration is found at this time allows one to estimate the probability with which it results from the randomisation process. We refer to the set of configurations obtained after randomisation for a certain time as the network ensemble, and the probabilities of each configuration in this ensemble as the network distribution.
Explicit randomisation has provided significant insight on the structural properties of real-world networks. Some examples include generating small-world networks\,\cite{watts1998collective}, the power-law degree distributions of preferential attachment mechanisms\,\cite{barabasi1999emergence} and the statistical analysis of social networks\,\cite{katz1957probability,holland1976local,rao1996markov,roberts2000simple}. However, small changes in the rules governing complex systems are known to bring about significant differences in their overall behaviour \cite{wolfram1984computation}. Small changes in randomisation steps that reflect these underlying rules can then also produce such differences. This can become a problem, for example, when attempting to reconstruct these rules in detail, or when different sets of rules bring about the same behaviour.

Distribution sampling methods, on the other hand, present a ``black-box'' version of this procedure. The network ensemble and distribution are defined instead of obtained, effectively selecting the set of all configurations considered to be possible results of the randomisation process and their probabilities. Randomisation is then achieved simply by sampling the ensemble according to the network distribution. While this method provides little insight on the underlying mechanism of the randomisation (i.e. what changes are allowed), the network distribution can be defined or deduced to reflect properties of the network to be randomised. In particular, the field of information theory has yielded a rigorous method to obtain network distributions based on statistical properties their ensembles are desired to have, establishing a formal framework for networks analogous to statistical mechanics\,\cite{park2004statistical,cimini2019statistical}. The method relies on analytically finding network distributions that maximise their Shannon entropy given specified constraints, that is, average values over the distribution. Constraints can then be chosen to reflect properties that are shared between the distribution on average and the original network to be randomised, such as the number of connections in the whole network. Meanwhile, the fact that these distributions maximise entropy allows them to be interpreted as being maximally random, or more precisely unbiased, with respect to properties that are not specified. In particular, it is known that distributions estimated from explicit randomisation at a large number of steps, i.e. when the network distribution becomes stationary, match the maximally random results from maximum entropy in some cases.  Samples drawn from this distribution can then be understood as randomised networks which, on average, retain the properties of the original network specified by constraints, but are maximally random otherwise. Applications are found in many areas, for example network construction\,\cite{garlaschelli2008maximum}, reconstruction from incomplete data\,\cite{squartini2011randomizingI,squartini2011randomizingII} and pattern detection\,\cite{squartini2017maximum} among others. However, as maximum entropy distributions are guaranteed to be unique, they cannot account for the variability needed to describe evolving systems.

While explicit randomisation has the advantage that it can account for the fundamentally dynamic (due to their interactions) nature of complex systems, the construction of randomisation steps does not count on foundations as rigorous and general as information theory. Additionally, the need to carry out large numbers of realisations of the process to obtain enough samples to measure statistical properties can quickly become a problem, for example in large networks. This is a problem that is easily avoided when the distribution of these samples is available, as is the direct result of maximum entropy-based methods. On the other hand, the dynamic aspect is not covered by most information theoretic applications to complex networks. This calls for an integrated information-theoretic method that both contemplates dynamic distributions of networks evolving by an underlying randomisation process and leads to a maximum entropy distribution in the stationary regime. 

Maximum caliber is the main tool of information theory to consider non-stationary processes\,\cite{jaynes1980minimum,ge2012markov,presse2013principles,dixit2018perspective,ghosh2020maximum}. Its main foundation is maximising Shannon entropy given certain constraints, providing the same interpretation of a distribution that is maximally unbiased with respect to properties that are not specified. While it is still guaranteed to produce a unique distribution, it contemplates evolution by studying probabilities of full trajectories in a dynamic process as opposed to individual states. Constraints then represent properties of the trajectories averaged over the distribution, for example the average number of connections in the whole evolution. As such, it is a strong contender for an information-theoretic method that captures the dynamic aspect of complex networks. However, literature on the application of maximum caliber to dynamic networks is not easy to find. Entropic dynamics\,\cite{caticha2011entropic,caticha2015entropic,pessoa2021entropic} might be considered as an exception, having been used in the study of dynamic networks from an information theoretic perspective. Nevertheless, its version of entropy is presented ad hoc and is therefore somewhat disconnected from both the stationary results of maximum entropy and the dynamic point of view of maximum caliber. 

The rest of the paper is structured as follows. In the next section, we introduce maximum caliber in the context of networks and establish how to obtain dynamic network configuration distributions for generic constraints. In \cref{app:maximum_entropy_production} we connect the formulation to entropic dynamics and show that it can be stated as a principle of maximum information entropy production, analogous to the thermodynamic theory for non-stationary processes\,\cite{martyushev2006maximum,martyushev2021maximum}. In the two sections that follow we consider specific constraint choices and randomisation steps, comparing the results of maximum caliber to distributions estimated from stochastic simulations to determine whether the information-theoretic method captures the explicit randomisation process.

\section{Maximum caliber networks}
\label{sec:maximum_caliber_networks}

A network is composed of a set of nodes, representing the components of a system, and a set of links, each associated to a pair of nodes and indicating that the pair can interact. In some cases, these interactions are directed from a source to a target, so links are associated with ordered pairs of nodes. These are known as directed networks, and examples include forces by particles on others and messages from people to their neighbours. In other cases, interactions are undirected, associating links to unordered pairs of nodes. These are called undirected networks and are the cases of the potential energy of pairs of particles or telephone lines between pairs of houses. For the results presented in \cref{sec:watts_strogatz,sec:degree_rewiring} we will consider undirected and directed networks respectively, but in both cases we will assume links are only associated to pairs of different nodes, i.e. we exclude self-interactions. It is then useful to refer to the pair formed from nodes $i$ and $j$ as the pair $ij$, with the understanding that this is an unordered pair $ij = ji$ if the network is undirected but an ordered one $ij \neq ji$ if $i \neq j$ and the network is directed. It is also useful to refer to $ij$ as a link when referring to the link between the pair $ij$, with the added understanding that in the cases considered this can only be a pair where nodes are different $i \neq j$. We will always explicitly state whether we refer to a link or a pair to keep the distinction clear. 

Given a network defined by a set of nodes and the links indicating which pairs can interact, a set $S$ of states that each link can take represents the different types of interactions in the system or their properties. It is useful to keep these states separate from the network itself as we can often represent the same system in different ways. For example, the average amount of energy $E$ that an individual gains or loses through interactions with each different member of an ecosystem could be described in a network through states in the real numbers, $E \in S = \mathbf{R}$. On the other hand, the same interactions might be classified as competitive ($E < 0$), cooperative ($E > 0$) or indifferent ($E = 0$). In this case each link can be associated to a state $s \in S = \{0,1,2\}$ representing one of the categories. As all links can be in the same states, it is not necessary to consider that the links themselves move between different pairs of nodes. The distinction between a network and its states then has the advantage of fixing the links in the network to specific pairs of nodes and relegating the dynamics of the network to changes in link states only. Here we will assume that each link can be found either in on or off states, indicating that the pair of nodes associated to the link is connected (on) or disconnected (off). The set of states is then $S = \{0,1\}$, with $1$ representing a link in a connected state and $0$ a disconnected one. Networks with states $S = \{0,1\}$ are known as binary, distinguishing them from cases where there are more than two link states such as the previous examples. The latter are known as weighted networks. The focus on binary networks emphasises that we will be studying the dynamics of the connection structure rather than dynamic processes unfolding on this structure. However, networks have found applications in describing both this binary structure and weighted connections\,\cite{farahani2019application}, and the framework of maximum caliber does not require specifying whether links are binary or not, suggesting that future work could make use of the methods presented here with weighted links.

Once a set of states $S$ is chosen, a particular collection of the states of each link in a network is called a network configuration. Note that if there are $N_L$ links in a network with states in $S$, a configuration is a vector of $N_L$ components, each taking a value in $S$. The set of all possible network configurations, here known as the configuration space $\mathcal{W}$, can then be understood as the set of all possible vectors with $N_L$ components in $S$, that is $\mathcal{W} = S^{N_L} = S \times S \times ... \times S$, the direct product of $S$ with itself $N_L$ times. As we consider that only pairs of different nodes are linked, the number of links is $N_L = N(N - 1)/2$ in the cases of undirected networks of $N$ nodes while $N_L = N(N - 1)$ for directed ones, defining the configuration spaces $S^{N_L}$ in each case. While network configurations are a vector with each element corresponding to a link, an adjacency matrix is an alternative representation that places focus on the nodes instead. It is a square matrix $W$ with $N$ rows and columns where each element $w_{ij}$  at row $i$ and column $j$ describes the interaction of the pair $ij$. In our case, if the pair $ij$ has an associated link, $w_{ij}$ represents the state of that link with $w_{ij} = 1$ for a connected state or $w_{ij} = 0$ for a disconnected one. If a link is not associated to the pair $ij$, the element of the adjacency matrix takes the value $w_{ij} = 0$, mimicking a disconnected link. All off-diagonal elements of the matrix are associated links, while diagonal elements are not. However, for undirected networks links associated to upper triangular elements of the matrix are the same as those associated to the lower triangular elements, meaning that the adjacency matrix is symmetric, i.e. $w_{ij} = w_{ji} ~ \forall ~ i,j$ as in this case any pair or link $ij$ is the same as the pair or link $ji$. Taking advantage of the notation introduced for links and pairs of nodes, we can also interpret the symbol $w_{ij}$ as the component corresponding to a link $ij$ in the vector representation of the network configuration, allowing us to easily switch between representations. For our choice of links among pairs of different nodes only, this technically involves restricting the indices of the matrix to all off-diagonal elements $i,j\neq i$ if the network is directed, or to the upper triangular elements $i,j>i$ if the network is undirected. To return to the adjacency matrix representation, we simply follow the procedure already described to construct the adjacency matrix. As the components of the adjacency matrix now share the symbol of the components of the network configuration, we will use the same symbol $W$ for the adjacency matrix and for the network configuration.

Given a configuration space $\mathcal{W}$, an evolving network can be described by a dynamic configuration $W(t) \in \mathcal{W} ~ \forall ~ t$. A sequence of $T+1$ network configurations $W_T = (W(0), W(1), W(2), \, ... \, , W(t),\, ...\, , W(T))$ then defines a $T$-step trajectory in the space $\mathcal{W}^{T+1}$, the direct product of the configuration space with itself $T+1$ times. We will now calculate the probability of these trajectories $P(W_T)$ using maximum caliber. Note that in the context of maximum caliber, the resulting distribution depends only on constraints chosen to represent properties of the evolution. Therefore any $W_T \in \mathcal{W}^{T+1}$ is a valid $T$-step trajectory, but depending on the constraints certain trajectories might be assigned a probability of zero. Constraints reflect properties of the evolution by requiring certain functions of the trajectory, known as constraint functions, to take specific average values, called constraint values, over the distribution of trajectories
\begin{equation}
    \sum_{W_T} F_n(W_T) P(W_T) = f_n \, .
    \label{eq:maxcal_constraints}
\end{equation}
For example, the average number of connections over the evolution is established by choosing a constraint function that sums over links $ij$ and times $t$, $F_1(W_T) = \sum_{0 \leq t \leq T} \sum_{ij} w_{ij}(t)$, and a constraint value $f_1$ which represents the numerical value of the average. Given a set of constraints, maximum caliber prescribes that, out of all trajectory distributions which have these average values, the one that is also maximally unbiased with respect to properties that are not imposed is the distribution that maximises the functional 
\begin{equation}
    S[P] = -\sum_{W_T} P(W_T) \ln(P(W_T)) \, .
    \label{eq:maxcal_entropy}
\end{equation}
The distribution that both satisfies \cref{eq:maxcal_constraints} and maximises \cref{eq:maxcal_entropy} is hereon referred to as the maximum caliber distribution. It can be found analytically by introducing Lagrange multipliers $\lambda_n$ for each of the constraints and maximising the Lagrangian
\begin{equation}
    \mathcal{L}[P] = S[P] + \sum_n \lambda_n \left( f_n - \sum_{W_T} F_n(W_T) P(W_T) \right) \, .
    \label{eq:maxcal_lagrangian}
\end{equation}
The introduction of the Lagrange multipliers allows one to ignore any dependence between the values of $P(W_T)$ for different $W_T$ in the maximisation. The distribution that achieves the supremum can then be obtained by deriving the Lagrangian with respect to a generic $P(W_T)$, equaling it to zero, and solving for $P(W_T)$ as in standard calculus.
\begin{equation}
    \begin{aligned}
    \frac{\partial \mathcal{L}[P]}{\partial P(W_T)} &= -1 - \ln(P(W_T)) - \sum_n \lambda_n F_n(W_T) = 0 \\
    \Rightarrow P(W_T) &= \exp \left( -1 -\sum_n \lambda_n F_n(W_T) \right)
    \end{aligned}
    \label{eq:lagrangian_max}
\end{equation}
While the result depends on the Lagrange multipliers $\lambda_n$, it can then be inserted into \cref{eq:maxcal_constraints} to solve for the dependence of the multipliers on each of the constraint values $f_m$, namely
\begin{equation}
    \sum_{W_T} F_m(W_T) \exp \left( -1 -\sum_n \lambda_n F_n(W_T) \right) = f_m \, .
    \label{eq:solve_multipliers}
\end{equation}
As there is one multiplier for each constraint, \cref{eq:solve_multipliers} constitutes a system of as many equations as unknowns. However, the non-linear character of this equation often calls for numerical or graphical methods to solve.

\subsection{Transition probabilities}

Because of the independence of the values $P(W_T)$ in the Lagrangian formalism, any relation between them in a particular maximum caliber distribution must arise from the constraints. The normalisation of the distribution $\sum_{W_T} P(W_T) = 1$ is such a relation. In most applications normalisation is ensured through a constraint function $F_o(W_T) = 1$ and constraint value $f_o = 1$, but here we will take a different approach.

Consider the $T$-step trajectory $W_T$ as a time-dependent $t$-step trajectory $W_t \in \mathcal{W}^{t+1}$ at time $t = T$. At each time $t>0$, a new configuration $W(t) \in \mathcal{W}$ is added to the trajectory from the previous time $W_{t-1}$, starting from $W_0 = W(0) \in \mathcal{W}$. This means that the updated trajectory can be constructed by pairing the trajectory from the previous time and its new configuration $W_t = (W_{t-1}, W(t))$. Additionally, the new trajectory space is the direct product of the one from the previous time and the configuration space $\mathcal{W}^{t+1} = \mathcal{W}^{t} \times \mathcal{W}$. This implies that any $t-1$-step trajectory can be paired with any network configuration giving a different $t$-step trajectory, and that all $t$-step trajectories are covered in this way. The joint probability of a $t-1$-step trajectory followed by a certain configuration is then the probability of the $t$-step trajectory they give rise to, $P(W_{t-1}, W) = P(W_t = (W_{t-1}, W))$. For a given trajectory $W_{t-1}$ at time $t-1$, different final configurations $W \in \mathcal{W}$ give rise to different $t$-step trajectories, all sharing $W_{t-1}$ as a history and differing only in this final state. If we sum over the probabilities of such $t$-step trajectories by summing over possible configurations $W \in \mathcal{W}$ as states at time $t = T$ after a particular $T-1$-step trajectory $W_{T-1}$, the result must agree with the probability of that $T-1$-step trajectory, relating the distribution of trajectories at time two successive times $T-1$ and $T$.
\begin{equation}
    P(W_{T-1}) = \sum_{W} P(W_{T-1}, W) \, .
    \label{eq:marginalisation_simple}
\end{equation}
Just as the normalisation of the probability distribution represents a relation among all probabilities of $T$-step trajectories, \cref{eq:marginalisation_simple} relates those with a shared history $W_{T-1}$. Additionally, summing \cref{eq:marginalisation_simple} over all $T-1$-step trajectories we have
\begin{equation}
    \sum_{W_{T-1}} P(W_{T-1}) = \sum_{W_{T-1}} \sum_{W} P(W_{T-1}, W) = \sum_{W_T} P(W_T) \, ,
    \label{eq:measure_conservation}
\end{equation}
where the second equality is due to the fact that the sum over all pairs $(W_{T-1}, W) \in \mathcal{W}^T \times \mathcal{W}$ covers all $W_T \in \mathcal{W}^{T+1}$. In particular, \cref{eq:measure_conservation} means that if the $T-1$-step distribution is normalised $\sum_{W_{T-1}} P(W_{T-1}) = 1$, then the $T$-step distribution is also normalised. Therefore if we ensure \cref{eq:marginalisation_simple} for all $T-1$-step trajectories through constraints, we simultaneously capture the dependence of the trajectory distribution on its past and guarantee that it remains normalised as it evolves.

In order to introduce constraints that ensure \cref{eq:marginalisation_simple}, consider an arbitrary $T$-step trajectory $W_T$ and $T-1$-step trajectory $W'_{T-1}$. Define a constraint function
\begin{equation}
    \delta_{W_{T-1}'}(W_T) := \begin{cases}
        1 & \text{if } W_{T} = (W_{T-1}', W) \\
        & ~~~~\text{for some $W \in \mathcal{W}$} \\
        0 & \text{otherwise}
    \end{cases}
\end{equation}
for each $W'_{T-1}$ which essentially takes a value of $1$ if $W_T$ is one of the $T$-step trajectories with $W'_{T-1}$ as a history. 
With these constraint functions, we construct the marginalisation constraints
\begin{equation}
    \sum_{W_T}  \delta_{ W_{T-1}' } ( W_T ) P( W_T ) = P(W_{T-1}') ~~ \forall ~ W_{T-1}'\, .
    \label{eq:maxcal_marginalisation}
\end{equation}
Note that the constraint function for each $W'_{T-1}$ ensures that the only trajectories that contribute to the sum are those that have a particular history $W_{T-1}'$ but different final configurations $W$, that is $\sum_{W_T} \delta_{W'_{T-1}}(W_T) P(W_T) = \sum_W P(W'_{T-1}, W)$.

Introducing marginalisation constraint functions explicitly into \cref{eq:lagrangian_max} along with corresponding multipliers $\lambda_{W_{T-1}'}$, but leaving room for constraints that are still unspecified, the $T$-step trajectory distribution becomes
\begin{equation}
    \begin{aligned}
    P(W_T) =& \exp \left( -1 -\sum_{W_{T-1}'} \lambda_{W_{T-1}'} \delta_{W_{T-1}'}(W_T) \right) \\ 
            & \times \exp \left( - \sum_n \lambda_n F_n(W_T) \right) \\
            & = \exp \left( - 1 - \lambda_{ W_{T-1} } - \sum_n \lambda_n F_n(W_T) \right) \, .
    \end{aligned}
    \label{eq:distribution_with_marginalisation_multihistory}
\end{equation}
The second equality is because there is only one $T-1$-step trajectory $W_{T-1} \in \mathcal{W}^{T}$ that is the history of a particular $T$-step trajectory $W_{T}$. In this sense, $W_{T-1}$ is not an arbitrary $T-1$-step trajectory but fixed by $W_T$. However, $W_T$ is arbitrary, so we can always choose it to be composed of an arbitrary $T-1$-step trajectory $W_{T-1}$ and any configuration. This is made explicit by expressing the $T$-step distribution as the joint distribution of $T-1$-step trajectories and final states, and introducing the notation $F(W_T) = F_n(W_{T-1}, W)$ for a trajectory $W_T = (W_{T-1}, W)$ 
\begin{equation}
    P(W_{T-1}, W) = \exp \left( -1 - \lambda_{W_{T-1}} - \sum_n \lambda_n F_n(W_{T-1}, W) \right)
    \label{eq:distribution_with_marginalisation}
\end{equation}
In order to remove the dependence on $\lambda_{W_{T-1}}$, we will introduce this expression in \cref{eq:maxcal_marginalisation} for $W'_{T-1} = W_{T-1}$. 
\begin{equation}
    \begin{aligned}
        P(W_{T-1}) = & \sum_{W_T} \delta_{W_{T-1}}(W_T) P(W_T) \\
        = & \sum_W P(W_{T-1}, W) \\
        = & \exp \bigg( - 1 -  \lambda_{W_{T-1}} \bigg) \times \\
        & \sum_{W} \exp \left(- \sum_n \lambda_n F_n(W_{T-1}, W) \right) \\
        \Rightarrow  \exp \left( -1 -\lambda_{W_{T-1}} \right)
    & = \frac{ P(W_{T-1}) }{ \sum_{W} \exp \left( - \sum_n \lambda_n F_n(W_{T-1}, W) \right)} \, .
    \end{aligned}
    \label{eq:solve_history}
\end{equation}

When introduced into \cref{eq:distribution_with_marginalisation}, this yields 
\begin{equation}
    P(W_{T-1}, W) = \frac{\exp \left( - \sum_n \lambda_n F_n(W_{T-1}, W) \right)}{\sum_{W} \exp \left( - \sum_n \lambda_n F_n(W_{T-1}, W) \right)} P(W_{T-1}) \, .
    \label{eq:maxcal_recursive}
\end{equation}

Note that the joint distribution in \cref{eq:maxcal_recursive} now marginalises to the $T-1$-step distribution by summing over final configurations. The joint distribution can then be divided by the $T-1$-step distribution to yield the conditional probability of a configuration $W(T)$ at time $T$ given a history $W_{T-1}$.
\begin{equation}
    \begin{aligned}
        P(W(T) | W_{T-1}) :&= \frac{P(W_{T-1}, W(T))}{P(W_{T-1})}\\
        &= \frac{\exp \left( - \sum_n \lambda_n F_n(W_{T-1}, W(T)) \right)}{\sum_{W} \exp \left( - \sum_n \lambda_n F_n(W_{T-1}, W) \right)} \, ,
    \end{aligned}
    \label{eq:maxcal_transitions} 
\end{equation}
We refer to the conditional probabilities as transition probabilities $M_T := P(W(T) | W_{T-1})$. This reflects the fact that at time $T$, a network trajectory $W_{T-1}$ will transition to the trajectory $W_T = (W_{T-1}, W(T))$ with probability $P(W(T) | W_{T-1})$. Transition probabilities depend entirely on imposed constraints (excluding marginalisation), and not on the trajectory distribution at the previous time as the joint probability does. They are then interpreted as reflecting the randomisation mechanism out of the context of a particular evolution, essentially playing the role of a randomisation step for maximum caliber.

If instead of summing over final configurations, we sum \cref{eq:maxcal_recursive} over $T-1$-step trajectories (i.e. possible histories), we can obtain the probability of a final configuration $W(T) \in \mathcal{W}$ at time $T$.
\begin{equation}
    P(W(T)) = \sum_{W_{T-1}} P(W_{T-1}, W(T)) = \sum_{W_{T-1}} M_T P(W_{T-1}) 
    \label{eq:dynamic_distribution}
\end{equation}

As for the history distribution $P(W_{T-1})$, recall that it was introduced through the constraint values of marginalisation constraints for each $T-1$-step trajectory. In the context of maximum caliber, constraint values are arbitrary numbers specified by the user of the method. We can therefore assume that the chosen constraint values match the maximum caliber trajectory distribution of $T-1$ step trajectories. Repeating the same process applied to $P(W_T)$ to obtain $P(W_{T-1})$ we see that the problem can be recursively reduced to the initial distribution $P(W_0) = P(W(0))$. We can then extend the validity of \cref{eq:dynamic_distribution} from $t = T$ to any $1 \leq t \leq T$, obtaining a dynamic distribution on a network ensemble $P(W(t))$. Samples of randomised networks can be drawn from this distribution at different times $t$, and its dynamics is determined by the choice of constraints beyond marginalisation. However, before showing that this distribution matches distributions estimated from stochastic simulations if constraints are chosen accordingly, we produce two useful results obtained by requiring the constraint functions to obey certain additional properties and explain how the comparison between simulations and analytical results is carried out.
 
\subsection{Markov processes}

While \cref{eq:maxcal_transitions} presents, in general, non-Markovian transition probabilities, the only dependence in the full history $W_{T-1}$ is through the constraint functions $F_n(W_{T-1}, W(T))$. This means that if all $F_n(W_{T-1}, W(T))$ depend on $W_{T-1}$ only through its final state $W(T-1)$, then transition probabilities depend only on two successive configurations $M_T = P(W(T) | W_{T-1}) = P(W(T)|W(T-1))$, and the process is Markovian. In this case, \cref{eq:dynamic_distribution} becomes the typical expression for updating the distribution of a Markovian process
\begin{equation}
    \begin{aligned}
    P(W(T)) &= \sum_{W_{T-1}} M_T P(W_{T-1})  \\
            &= \sum_{W(T-1)} M_T \sum_{W_{T-2}} P(W_{T-2}, W(T-1)) \\
            &= \sum_{W(T-1)} M_T P(W(T-1)) \, .
    \end{aligned}
    \label{eq:markov_dynamics}
\end{equation}
The second line is obtained by regarding $T-1$-step trajectories $W_{T-1}$ as pairs of $T-2$-step trajectories $W_{T-2}$ and final states $W(T-1)$ just as was done for $T$-step trajectories. Because $M_T$ is assumed not to depend on $W_{T-2}$, only on $W(T-1)$ and $W(T)$, it can be factored out of the sum over $T-2$-step trajectories. In the third line, the sum over $T-2$-step trajectories marginalises the $T-1$-step trajectory distribution to the distribution of configurations at time $T-1$ in the same way that the $T$-step trajectory is marginalised to the distribution of configurations at time $T$ in \cref{eq:dynamic_distribution}. In previous literature it has been established that Markov processes emerge when constraints specify the state of the system at individual instants in time\,\cite{ge2012markov,davis2015hamiltonian} (essentially each $F_n(W_T)$ depends on a single $W(t)$ in $W_T$), so this represents an extension of that condition. This is discussed in greater detail in \cref{app:maximum_caliber_and_markov}, where it is shown how some constraints placed on the entire sequence of states (as is common in maximum caliber) can hide constraints that specify the state of the system at each instant. 

\subsection{Independent links}

Consider now the possibility that constraint functions in the transition probabilities can be expressed as a linear combination of functions, each depending on the sequence of states composing the trajectory of a particular link $ij$ in the network $w_{ij}^{T} := (w_{ij}(0), w_{ij}(1), w_{ij}(2), ..., w_{ij}(T))$,
\begin{equation}
    \sum_n \lambda_n F_n(W_T) = \sum_{ij} \sum_m \lambda_{ij}^{m} G^{m}_{ij}(w_{ij}^T) \, .
    \label{eq:constraint_decomposition}
\end{equation}

With constraints in the form of \cref{eq:constraint_decomposition}, and introducing $G^m_{ij}(w_{ij}^T) = G^m_{ij}(w_{ij}^{T-1}, w_{ij}(T))$ for a link trajectory $w_{ij}^T = (w_{ij}^{T-1}, w_{ij}(T))$ composed of the states of $w_{ij}^{T-1}$ until time $T-1$ and a final state $w_{ij}(T)$, \cref{eq:maxcal_transitions} becomes
\begin{equation}
    \begin{aligned}
    M_T &= \frac{\exp \left( -\sum_{ij} \sum_m \lambda_{ij}^{m} G^{m}_{ij}(w_{ij}^{T-1}, w_{ij}(T)) \right)}{ \sum_{W} \exp \left( - \sum_{ij} \sum_m \lambda_{ij}^{m} G^{m}_{ij}(w_{ij}^{T-1}, w_{ij}) \right) }  \\ 
        &= \frac{ \prod_{ij} \exp \left( - \sum_m \lambda_{ij}^{m} G^{m}_{ij}(w_{ij}^{T-1}, w_{ij}(T)) \right)}{ \sum_{kl} \sum_{w_{kl}} \prod_{ij} \exp \left( - \sum_m \lambda_{ij}^{m} G^{m}_{ij}(w_{ij}^{T-1}, w_{ij}) \right) }  \\ 
        &= \frac{ \prod_{ij} \exp \left( -\sum_m \lambda_{ij}^{m} G^{m}_{ij}(w_{ij}^{T-1}, w_{ij}(T)) \right) }{ \prod_{ij} \sum_{w_{ij}} \exp \left( -\sum_m \lambda_{ij}^{m} G^{m}_{ij}(w_{ij}^{T-1}, w_{ij}) \right) } \, .
    \end{aligned}
    \label{eq:independent_transitions}
\end{equation}
The denominator in the second line of \cref{eq:independent_transitions} is obtained by decomposing the sum over all configurations $W$ into sums over the states $w_{kl}$ of each link $kl$. In the third the denominator results from carrying out the sum over states $w_{ij}$ of a particular link $ij$ after factoring out the products that correspond to all other links $kl \neq ij$ as these do not depend on $w_{ij}$. Thus, \cref{eq:independent_transitions} shows that under constraints of the form of \cref{eq:constraint_decomposition}, the transitions of the network can be written as a product of transition probabilities $P_{ij}(w_{ij}(T) | w_{ij}^{T-1})$ corresponding to the trajectory of each link, 
\begin{equation}
    \begin{aligned}
    P_{ij}(w_{ij}(T) | w_{ij}^{T-1}) &= \frac{ \exp \left( - \sum_m \lambda_{ij}^{m} G^{m}_{ij}(w_{ij}^{T-1}, w_{ij}(T)) \right) }{ \sum_{w_{ij}} \exp \left( -\sum_m \lambda_{ij}^{m} G^{m}_{ij}(w_{ij}^{T-1}, w_{ij}) \right) } \\
    M_T &= \prod_{ij} P_{ij}(w_{ij}(T) | w_{ij}^{T-1}) 
    \end{aligned}
    \label{eq:transition_product}
\end{equation}

If the transitions take the form of \cref{eq:transition_product} and the distribution of network trajectories of length $T-1$ is a product of independent link trajectory distributions $P_{ij}(w_{ij}^{T-1})$, that is $P(W_{T-1}) = \prod_{ij} P_{ij}(w_{ij}^{T-1})$ then so will the distribution of trajectories of $T$ steps $P(W_T) = \prod_{ij} P_{ij}(w_{ij}(T) | w_{ij}^{T-1}) P_{ij}(w_{ij}^{T-1}) = \prod_{ij} P_{ij}(w_{ij}^{T})$. Marginalising this expression over state trajectories of all except a particular link, we find its trajectory can be updated independently from others. For any link $ij$, the probability of a trajectory $w_{ij}^{T} = (w_{ij}^{T-1}, w_{ij}(T))$ is given by
\begin{equation}
    P_{ij}(w_{ij}^{T}) = P_{ij}(w_{ij}(T) | w_{ij}^{T-1}) P_{ij}(w_{ij}^{T-1}) ~~ \forall ~ ij \, .
\end{equation}
Additionally, whenever the network trajectory distribution is a product of independent link trajectory distributions, then the network distribution $P(W(T))$ is a product of individual link probabilities $P_{ij}(w_{ij}(T))$ since, following the same logic as the second and third lines of \cref{eq:independent_transitions},
\begin{equation}
    \begin{aligned}
    P(W(T)) &= \sum_{W_{T-1}} P(W_T) \\
            &= \sum_{kl} \sum_{w_{kl}^{T-1}} \prod_{ij} P_{ij}(w_{ij}^T) \\ 
            &= \prod_{ij} \sum_{w_{ij}^{T-1}} P_{ij}(w_{ij}^T) \\
            &= \prod_{ij} P_{ij}(w_{ij}(T)) \, .
    \end{aligned}
\end{equation}
Due to the recursive nature of the results on link independence, the ability to factorise network distributions into independent link probabilities can be traced back to the choice of initial network distribution $P(W(0))$.

\subsection{Comparing stochastic simulations and maximum caliber} 

To assess whether maximum caliber can capture the dynamic distribution of an ensemble of networks undergoing a stochastic process, we consider an ensemble of $R$ network configurations $\{W^1(0), W^2(0), ..., W^r(0), ..., W^{R}(0)\}$ drawn from a distribution $P(W(0))$. A configuration is sampled from a binary network distribution with independent links $P(W(0)) = \prod_{ij} P(w_{ij}(0))$, as is assumed to be the case from here on, by starting with a fully disconnected network with the same nodes and links as the network distribution. Next, a random number $x_{ij}$ uniformly distributed between $0$ and $1$ is drawn for each different link $ij$. The sampled configuration is then constructed by setting the state of the link $ij$ to connected if $x_{ij} < P(w_{ij}(0) = 1)$. The distribution $P(W(0))$ will be used as the initial condition for maximum caliber while configurations in the ensemble are initial conditions for different realisations of the randomisation process. This ensures that, at least initially, the distribution of maximum caliber represents the explicitly randomised ensemble. Note that if the initial distribution is of the type $P(W(0)) = \delta_{W(0),V} = \prod_{ij} \delta_{w_{ij}(0),v_{ij}}$, where $v_{ij}$ is the state of the link $ij$ in a configuration $V$, then all samples drawn from the distribution are $V$ and therefore all randomisation trajectories start from the same configuration.

Given the ensemble and the distribution it is drawn from, we choose a set of constraints for maximum caliber and a randomisation step for stochastic simulations. The constraints of maximum caliber allow to calculate the transitions of the initial probability distribution $P(W(0))$ to $P(W(1))$ while the randomisation step is applied to each configuration $W^r(0)$ drawn from the initial distribution of maximum caliber obtaining once-randomised samples $W^r(1)$. 

Once samples have been obtained by the first step of explicit randomisation, we can estimate their distribution and compare it to the one updated by maximum caliber to find out whether they are the same. Both for the estimation and comparison of binary networks distributions of independent links it is useful to define a probability matrix $P$ with the same shape as the adjacency matrix and values $p_{ij}$ indicating the probability that the element $w_{ij}$ of the network adjacency matrix takes a value of $1$. If the element corresponds to a linked pair of nodes, this is the probability of the link in a connected state $p_{ij} = P_{ij}(w_{ij} = 1)$. If the element does not have an associated link, this probability is $p_{ij} = 0$. The probability matrix is enough to fully capture binary network distributions as a connected link $ij$ has a probability $P_{ij}(w_{ij} = 1) = p_{ij}$ by definition and a disconnected one $P_{ij}(w_{ij} = 0) = 1-p_{ij}$ because of normalisation. Thus probability matrices facilitate the estimation of sample distributions as it is easier to measure the probability matrix than the distribution. From network samples $W^r$, the estimated probability that a given element of the adjacency matrix takes the value $w_{ij} = 1$ is the average value of that element over samples, $p_{ij} = \sum_r w^{r}_{ij} / R$ and therefore the estimated probability matrix is simply the average adjacency matrix of the samples $P = \sum_r W^{r}/R$. From maximum caliber, the values of the probability matrix result by definition from the network distribution.  

\begin{figure}[h!]
    \centering
    \includegraphics[width=0.9\linewidth]{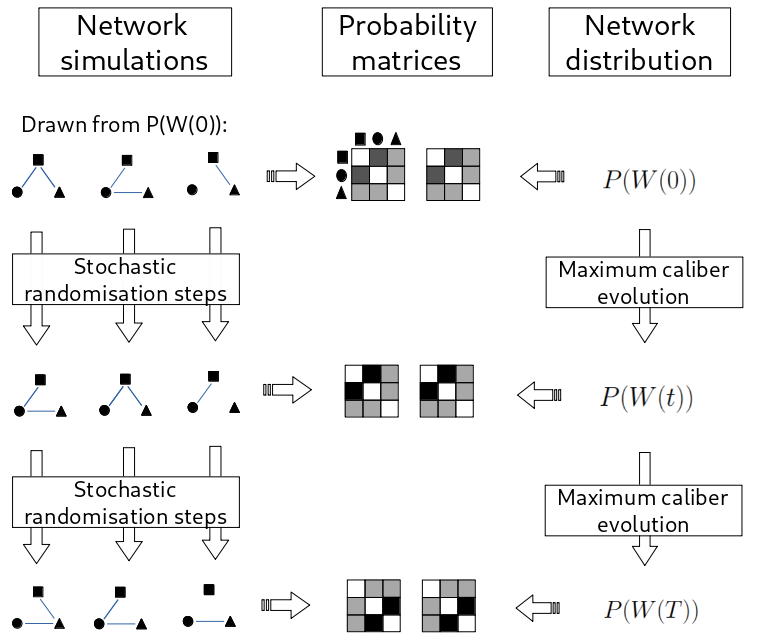}
    \caption{On the right side, an initial distribution is evolved by maximum caliber for $T$ steps. On the left side, networks drawn from the initial distribution used by maximum caliber are evolved according to a stochastic randomisation simulation for $T$ steps. At each step, maximum caliber produces a probability matrix and the network simulations estimate one, which are the same if the processes agree.}
    \label{fig:comparison}
\end{figure}

After the probability matrices have been obtained by each method at the first step, the transitions of maximum caliber once again update the maximum caliber distribution, and a randomisation step is applied to each once-rewired network sample $W^r(1)$, obtaining a twice rewired $W^r(2)$. The estimation and calculation of the probability matrices is repeated, allowing for a new iteration. After $T$ repetitions, the methods are compared by examining if the estimated and calculated probability matrices are the same at each time up to $T$. This concludes the description of the procedure to test whether maximum caliber distributions can capture stochastic network evolution, represented graphically in \cref{fig:comparison}.

\begin{figure}[h!]
    \centering
    \includegraphics[width=\linewidth]{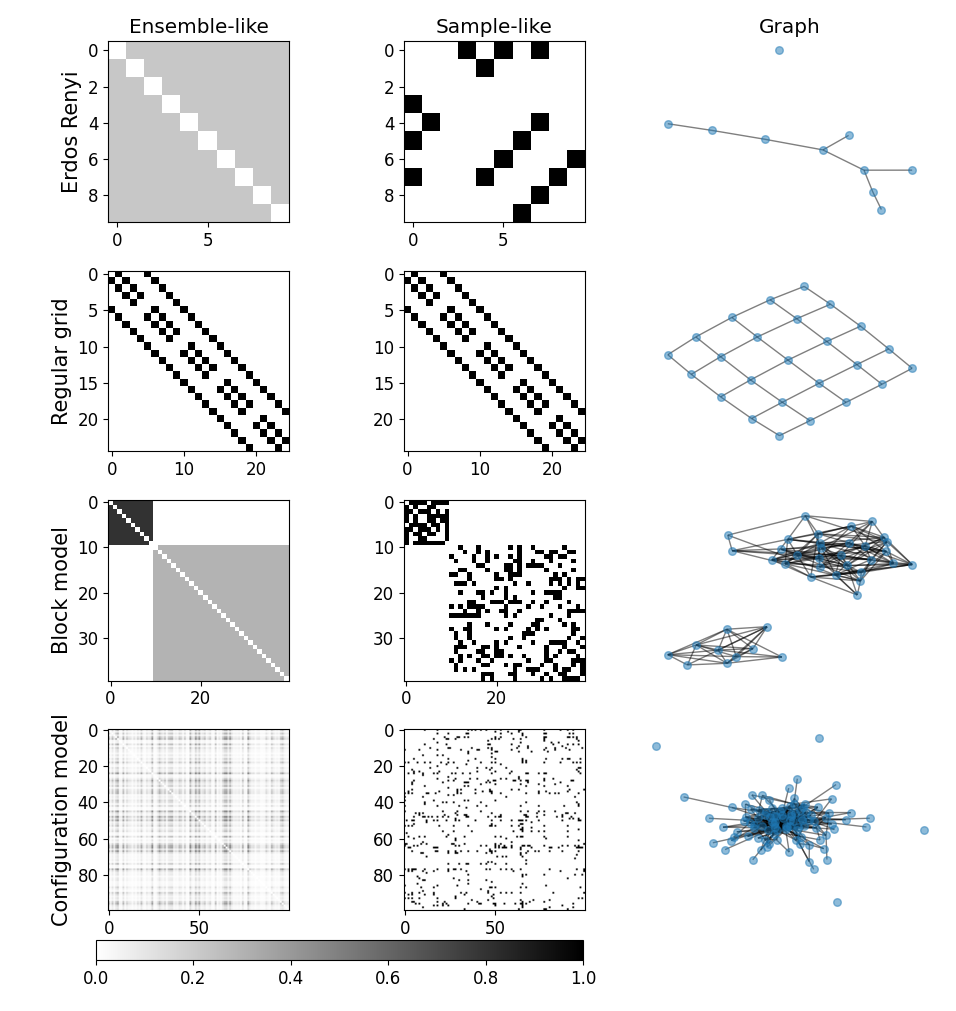}
    \caption{Initial conditions for comparison between analytical results from maximum caliber and stochastic simulations. In each row, from left to right: Ensemble-like initial probability matrices, sample-like initial probability matrices, and graph representations of the samples. From top to bottom, an Erd\"{o}s R\'{e}nyi network ensemble distribution (ER), a 2-dimensional regular grid (RG), a block model (BM), and a configuration model (CM).}
    \label{fig:initial_conditions}
\end{figure}

Throughout the cases considered in the following sections, we consider eight initial conditions. Each one defines an initial distribution for maximum caliber and is therefore also used to sample initial configurations for realisations of explicit randomisation. The first four of these are what we refer to as ensemble-like initial conditions, shown as heatmaps of their probability matrices in the first column from left to right of \cref{fig:initial_conditions}, and consist of
\begin{itemize}
    \item[ER -] $10$ node Erd\"{o}s-R\'{e}nyi: the maximum entropy network ensemble distribution resulting from constraining the total amount of connections in a binary network. The resulting connection probability of any link is the same value, in this case $2/9$.
    \item[RG -] $25$ node regular grid: neighbouring nodes on a two-dimensional $5 \times 5$ grid are connected with probability $1$ and non-neighbouring ones are connected with probability $0$.
    \item[BM -] $40$ node block model: two blocks of $10$ and $30$ nodes with uniform probabilities of $0.8$ and $0.3$ between pairs of different nodes in each block respectively and a connection probability of $0$ for pairs of nodes belonging to different blocks.
    \item[CM -] $100$ node binary configuration model: the maximum entropy network ensemble distribution resulting from constraining the degree sequence of a network, the number of connections of each node. The resulting probability of any particular node pair is $p_{ij} = 1/(1 + \exp(\lambda_i + \lambda_j))$. In this case, the values of $\lambda$ were drawn independently from a uniform distribution between $0$ and $3$ instead of resulting from a particular degree sequence for simplicity.
\end{itemize}
The other four initial conditions are determined by drawing a configuration $V$ from each of the aforementioned network distributions and defining the distributions $\delta_{W(0),V}$ for each one. We refer to these as sample-like initial conditions, and note that the probability matrices of these distributions, shown in the second column of \cref{fig:initial_conditions}, are equal to the adjacency matrices of the configurations $V$ that give rise to them. The graph of each is shown in the third column of the same figure. For each initial condition, the number of nodes $N$ in the network determines the total of steps in the evolution, which is $20 N$ in all cases, and the number of realisations to estimate explicitly randomised distributions, $10 N$ except for the configuration model ensemble and sample like initial condition, for which $N$ samples are used.

In the next two sections, we consider particular cases of randomisation steps and maximum caliber constrained dynamics. In \cref{sec:watts_strogatz} these are Watts-Strogatz rewiring\,\cite{watts1998collective} and the conservation of the number of connections, along with some variations. In \cref{sec:degree_rewiring} they are degree-preserving rewiring\,\cite{katz1957probability,holland1976local,rao1996markov,roberts2000simple} and the conservation of the degree sequence. From the initial conditions described, we compare the distribution of each method over time, showing that maximum caliber dynamics captures the evolving distribution of explicitly randomised network ensembles. 

\section{Maximum entropy Watts-Strogatz rewiring}
\label{sec:watts_strogatz}

A single Watts-Strogatz rewiring step consists of choosing, uniformly and at random, one among the $L$ connected links of a binary network and one among the $N_L - L$ disconnected links (with $N_L$ the total number of links). With a replacement probability $p$, the states of the links are exchanged, with the connected link becoming disconnected and vice-versa. With probability $1-p$, no change is made. Note that, on average, a given value of $p$ requires $1/p$ times the number of steps to achieve the same randomisation. Similarly, directed networks where the link $ij$ is considered different from $ji$ require randomising twice the number of links. We will therefore focus on the case where the replacement probability is $p=1$ and undirected links among pairs of different nodes. 

By construction, Watts-Strogatz rewiring conserves the number of connections $L$ in a network throughout its application. As constraints capture characteristic traits of the network trajectory evolution and due to the recursive nature of maximum caliber, we impose the conservation of the average number of connections in the ensemble with respect to the previous step,
\begin{equation}
    \sum_{W_T} \sum_{i,j>i} \left( w_{ij}(T) - w_{ij}(T-1) \right) P(W_T) = 0 \, .
    \label{eq:link_conservation}
\end{equation}
Note that the sum over links $ij$ is carried out by summing only the upper triangular adjacency matrix elements $i,j>i$.

Additionally, another constraint is needed. Watts-Strogatz rewiring as described defines a single step by making exactly two changes in the configuration of the network, one link disconnecting and another connecting. The number of such changes can be measured by counting the links that change states, regardless of whether they change from $w_{ij}(T-1) = 0$ to $w_{ij}(T) = 1$ or $w_{ij}(T-1) = 1$ to $w_{ij}(T) = 0$. For this the constraint is
\begin{equation}
    \sum_{W_T} \sum_{i,j>i} | w_{ij}(T) - w_{ij}(T-1) | P(W_T) = 2 \, .
    \label{eq:successive_difference}
\end{equation}
Note that the process defined by these constraints is Markov as both depend only on the two last states $W(T-1)$ and $W(T)$ in the network trajectory. Also, introducing multipliers $\alpha$ and $\beta$ for \cref{eq:link_conservation} and \cref{eq:successive_difference} respectively, we have
\begin{equation}
    \begin{aligned}
    & \sum_n \lambda_n F_n(W_T) \\
    & ~~ = \sum_{i,j>i} \alpha \left( w_{ij}(T) - w_{ij}(T-1) \right) + \beta | w_{ij}(T) - w_{ij}(T-1) |
    \end{aligned} 
\end{equation}
meaning that \cref{eq:constraint_decomposition} is valid with $\lambda^0_{ij} = \alpha$, $\lambda^{1}_{ij} = \beta$, $G^{0}_{ij}(w^T_{ij}) = w_{ij}(T) - w_{ij}(T-1)$ and $G^{1}_{ij}(w^T_{ij}) = |w_{ij}(T) - w_{ij}(T-1)|$. By the results from \cref{sec:maximum_caliber_networks} the transition probability $M_T$ is a product of independent Markovian link transitions
\begin{equation}
    \begin{aligned}
    & P_{ij}(w_{ij}(T) | w_{ij}^{T-1}) = P_{ij}(w_{ij}(T) | w_{ij}(T-1)) \\
    &= \frac{\exp \left( -\alpha \left( w_{ij}(T) - w_{ij}(T-1) \right) - \beta | w_{ij}(T) - w_{ij}(T-1) | \right)}{\sum_{w_{ij}} \exp \left( - \alpha \left( w_{ij} - w_{ij}(T-1) \right) - \beta | w_{ij} - w_{ij}(T-1) |\right)}
    \end{aligned}
\end{equation}
As the network links are binary, these link transitions define the annihilation probabilities
\begin{equation}
    \begin{aligned}
    a_{ij} :&= P_{ij}(w_{ij}(T) = 0 | w_{ij}(T-1) = 1) \\
            &= \frac{\exp \left( \alpha - \beta \right)}{\sum_{w_{ij} = 0}^{1} \exp \left( -\alpha \left( w_{ij} - 1 \right) - \beta | w_{ij} - 1 |\right)} \\
            &= \frac{1}{1 + \exp \left( - \alpha + \beta  \right)}
    \end{aligned}
\end{equation}
and creation probabilities
\begin{equation}
    \begin{aligned}
    c_{ij} :&= P_{ij}(w_{ij}(T) = 1 | w_{ij}(T-1) = 0) \\
            &= \frac{\exp \left( - \alpha - \beta \right)}{\sum_{w_{ij} = 0}^{1} \exp \left( - \alpha w_{ij} - \beta | w_{ij} | \right)} \\
            &= \frac{1}{1 + \exp \left( \alpha + \beta  \right)} \, .
    \end{aligned}
\end{equation}
The link transitions can define link-specific transition matrices
\begin{equation}
    \begin{aligned}
        m_{ij} :&= P_{ij}(w_{ij}(T) | w_{ij}(T-1)) \\
        &= \begin{pmatrix}
            1 - c_{ij} & a_{ij} \\
            c_{ij} & 1 - a_{ij} 
        \end{pmatrix} \\
        &= \begin{pmatrix}
        \frac{1}{1 + \exp( -\alpha - \beta))} & \frac{1}{1 + \exp( - \alpha + \beta)} \\
        \frac{1}{1 + \exp( \alpha + \beta)} & \frac{1}{1 + \exp( \alpha - \beta))}            
        \end{pmatrix}
    \end{aligned}
\end{equation}
where the entries with values $1-c_{ij}$ and $1-a_{ij}$ result from marginalisation. As the multipliers are independent of each specific link $ij$, we find that the link transition matrix is the same for each link in the network, that is $c_{ij} = c = 1/(1 + \exp(\alpha + \beta))$ and $a_{ij} = a = 1/(1 + \exp (- \alpha + \beta))$ for all $ij$. The values of these probabilities can be found analytically by imposing constraints \cref{eq:link_conservation,eq:successive_difference},
\begin{equation}
    \begin{aligned}
        0 &= \sum_{W_T} \sum_{i,j>i} \left( w_{ij}(T) - w_{ij}(T-1) \right) P(W_T) \\
        &= \sum_{W_T} \sum_{i,j>i} \left( w_{ij}(T) - w_{ij}(T-1) \right) \prod_{i,j>i} P_{ij}(w_{ij}^T) \\
        &= \sum_{i,j>i} \sum_{ \substack{w_{ij}(T) \\ w_{ij}(T-1)}} \left( w_{ij}(T) - w_{ij}(T-1) \right) m_{ij} P_{ij}(w_{ij}(T-1)) \\
        &= \sum_{i,j>i} c (1-p_{ij}(T-1)) - a p_{ij}(T-1) \\
        &= (N(N-1)/2 - L) c - a L
    \end{aligned}
    \label{eq:impose_link_conservations}
\end{equation}
and
\begin{equation}
    \begin{aligned}
        2 &= \sum_{W_T} \sum_{i,j>i} | w_{ij}(T) - w_{ij}(T-1) | P(W_T) \\
        &= \sum_{W_T} \sum_{i,j>i} | w_{ij}(T) - w_{ij}(T-1) | \prod_{i,j>i} P_{ij}(w_{ij}^T) \\
        &= \sum_{i,j>i} \sum_{ \substack{w_{ij}(T) \\ w_{ij}(T-1)} } | w_{ij}(T) - w_{ij}(T-1) | m_{ij} P_{ij}(w_{ij}(T-1)) \\
        &= \sum_{i,j>i} c (1-p_{ij}(T-1)) + a p_{ij}(T-1) \\
        &= (N(N-1)/2 - L) c + a L \, .
    \end{aligned}
    \label{eq:impose_successive_difference}
\end{equation}
In both cases, the second line is obtained by expanding the network trajectory distribution into a product of independent link trajectory distributions. The third results from summing over the trajectories of all links in the product that are not multiplied by the corresponding $G_{ij}^m(w_{ij})$, all of which are normalised. The last line explicitly introduces the value $L = \sum_{i,j>i} p_{ij}(T-1)$ of the average number of connections of the network distribution $P(W(T-1))$. The creation and annihilation probabilities thus result in
\begin{equation}
    c = \frac{1}{N(N-1)/2 - L} ~~~~~~~~ a = \frac{1}{L}
\end{equation}

\begin{figure}[h!]
    \centering
    \includegraphics[width=\linewidth]{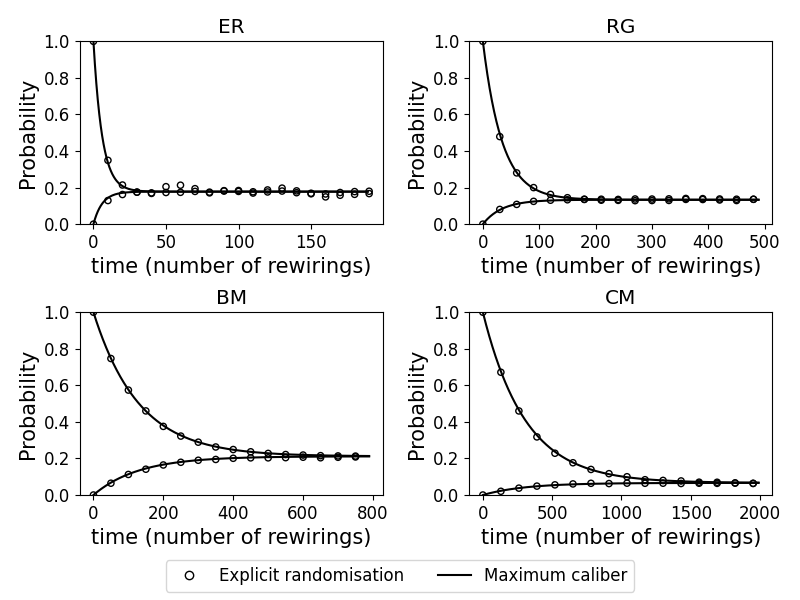}
    \caption{Probability of an initially connected link (starting from a probability of $1$) and an initially disconnected one (starting with a probability of $0$) according to simulations in circular markers and maximum entropy in full lines. Initial conditions correspond to sample-like initial conditions, shown in the second and third columns (from left to right) of \cref{fig:initial_conditions}.}
    \label{fig:WS_con_sample_rewirings}
\end{figure}

\begin{figure}
    \centering
    \includegraphics[width=\linewidth]{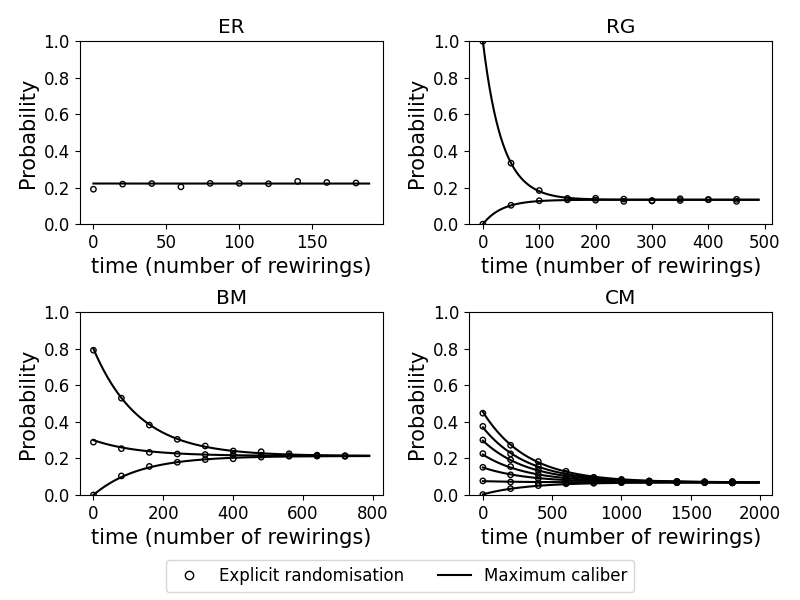}
    \caption{From the ensemble-like initial conditions shown in \cref{fig:initial_conditions}, each figure shows the probability of specific pairs of nodes being connected over time. Results from explicit randomisation in circular markers maximum caliber in full lines.}
    \label{fig:WS_con_ensemble_rewirings}
\end{figure}

Having described the explicit randomisation process and obtained the transition probabilities according to maximum caliber, we will now compare the evolution of the resulting network distributions from the different initial conditions represented in \cref{fig:initial_conditions}. For this, we choose a subset of the links for each initial condition and show the evolution of their connection probabilities according to the value of their respective entries in the probability matrix of explicit randomisation and maximum caliber. In \cref{fig:WS_con_sample_rewirings} we show the evolution of connection probabilities starting from sample-like initial conditions while \cref{fig:WS_con_ensemble_rewirings} are from ensemble-like initial conditions. Values obtained from explicit randomisation are shown in circular markers while full lines represent maximum caliber. 

From sample-like initial conditions as those shown in \cref{fig:WS_con_sample_rewirings}, connection probability values all start at $1$ or $0$, expected as the initial probability matrix matches the adjacency matrix of each sample. We therefore show the evolution of two links, an initially connected one starting with a connection probability of $1$ and an initially disconnected one starting with a connection probability of $0$. We have verified that other links present the same evolution (for each initial condition), as expected by the creation and annihilation probabilities not depending on specific links $ij$. Additionally, the evolution of all entries converges to the same probability at long times (for each network), matching the equilibrium distribution of Watts-Strogatz rewiring. This corresponds to the Erd\"{o}s-R\'{e}nyi distribution, the maximum entropy distribution constrained by the number of connections in the network.
 
Evolution from ensemble-like initial conditions are presented in \cref{fig:WS_con_ensemble_rewirings}. In Erd\"{o}s-R\'{e}nyi initial conditions (ER) all links have the same initial probability of being connected, and therefore the evolution of a single entry in the probability matrix is representative of the whole network ensemble. Additionally, because the Erd\"{o}s-R\'{e}nyi random graph is the equilibrium state of the rewiring process and the maximum entropy distribution, the network distribution is unchanged by the rewiring process. For the regular grid (RG), neighbouring nodes have a probability of $1$ of being connected while all others have probability $0$, so the observed evolution is identical to the one in \cref{fig:WS_con_sample_rewirings}. The block model (BM) takes three possible connection probabilities corresponding to links within the small block, large block, or between them, resulting in three different evolutions from the three possible initial probability values. Finally, the configuration model (CM) practically takes a continuum of initial probability values, so we have chosen $7$ links with approximately evenly spaced initial probabilities to compare their behaviour over time.

In both the case of sample-like and ensemble-like initial conditions, the evolution according to maximum caliber matches that of explicit randomisation with high accuracy. The constraints that define the analytic dynamics reflect properties of the underlying randomisation process and the constraints of the equilibrium maximum entropy distribution. This indicates that the method is well suited to replace realisations of dynamical processes on networks, yielding the distribution of trajectories based on constraints and an initial condition, deriving the evolution analogously to how traditional maximum entropy methods yield the distribution of equilibrium states. 

\subsection{Variation of the average number of connections}

The conservation of the average number of connections in \cref{eq:link_conservation} can be extended to a more general case. Notice that the constraint function on the left-hand side of the equation establishes what property of the dynamics is imposed, in this case the change in the average number of connections between time $T-1$ and $T$, while the constraint value on the right-hand side sets the numerical value of said property. In the particular case presented, this value is $0$, reflecting the conservation of the average number of connections throughout the evolution. If, on the other hand, the value is different from $0$, the constraint function still represents the change in the average number of connections, but the value no longer indicates a conservation law. Nevertheless, no mention of the constraint values is made until the Lagrange multipliers are found by setting the functional form of transitions $M_T$ into the constraints. The only difference between a conservation law of a certain property and the case where the change of the same property is specified but non-zero is in this last step. 

\begin{figure}[h!]
    \centering
    \includegraphics[width=\linewidth]{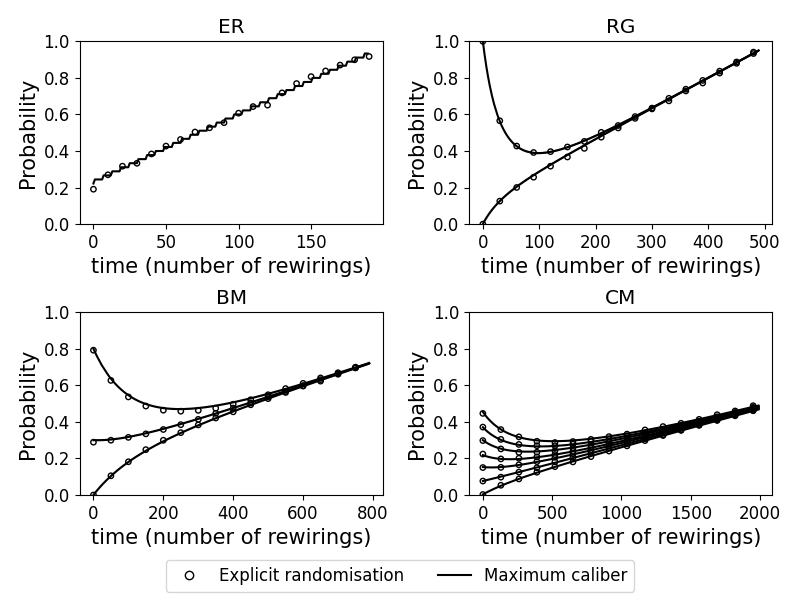}
    \caption{Evolution of connection probabilities between selected links under the linear increase of the average number of connections from ensemble-like initial conditions. Circular markers show results from explicit randomisation while full lines show the evolution according to maximum caliber.}
    \label{fig:WS_var_evolution_ensemble_entries}
\end{figure}

To test maximum caliber constraints beyond conservation laws, we consider two variations of the results already presented in this section. The first considers an explicit randomisation process identical to the described Watts-Strogatz rewiring with the modification that, in addition to the exchange of states of a connected and disconnected link, every $\tau$ steps a randomly chosen disconnected link is connected. For maximum caliber, the change of the average number of connections can be described by $\Delta_c(T)=1$ if $T \text{ mod } \tau = 0$ and $\Delta_c(T) = 0$ otherwise, converting \cref{eq:link_conservation} to
\begin{equation}
    \sum_{W_T} \sum_{i,j>i} \left( w_{ij}(T) - w_{ij}(T-1) \right) P(W_T) = \Delta_c(T) \, ,
    \label{eq:link_variation}
\end{equation}
and the number of changes made in the network configuration, \cref{eq:successive_difference}, becomes
\begin{equation}
    \sum_{W_T} \sum_{i,j>i} | w_{ij}(T) - w_{ij}(T-1) | P(W_T) = 2 + \Delta_c(T) \, .
    \label{eq:successive_difference_variation}
\end{equation}
Because the constraint functions are still the same, the network trajectory transition matrix $M_T$ is a product of independent link transition matrices $m_{ij}$ with annihilation and creation probabilities $a_{ij} = a$ and $c_{ij} = c$ that are the same for every link. Their values can be found by introducing the transitions into \cref{eq:link_variation,eq:successive_difference_variation}. Following the same steps as in \cref{eq:impose_link_conservations,eq:impose_successive_difference} we have
\begin{equation}
    \begin{aligned}
        \Delta_c(T) &= (N(N-1)/2 - L) c - a L \\
        2 + \Delta_c(T) &= (N(N-1)/2 - L) c + a L
    \end{aligned}
\end{equation}
which results in
\begin{equation}
    c = \frac{1 + \Delta_c(T)}{N(N-1)/2 - L} ~~~~~~~~ a = \frac{1}{L} \, .
\end{equation}
In \cref{fig:WS_var_evolution_ensemble_entries} we show the evolution of the same probability matrix values presented in \cref{fig:WS_con_ensemble_rewirings} starting from ensemble-like initial conditions with varying average number of connections according to simulations and maximum caliber. The values of $\tau$ are $\tau_{\text{ER}} = 6$, $\tau_{\text{RG}} = 2$, $\tau_{\text{BM}} = 2$ and $\tau_{\text{CM}} = 1$ for the Erd\"{o}s-R\'{e}nyi, regular grid, block model, and configuration model initial conditions respectively. As in the case of conserved average number of connections, probabilities obtained from maximum caliber and simulations are found to match.

\begin{figure}[h!]
    \centering
    \includegraphics[width=\linewidth]{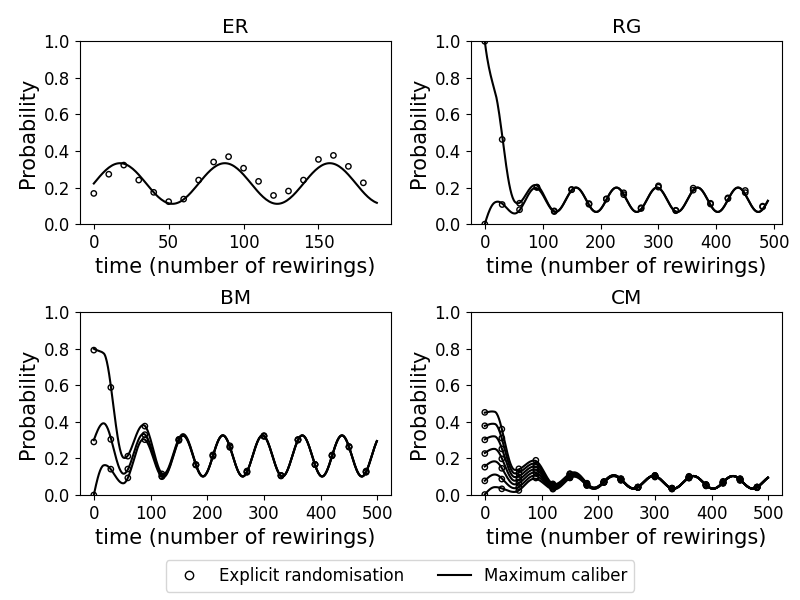}
    \caption{Evolution of link probabilities between selected pairs of nodes under an oscillating number of links from ensemble-like initial conditions. Circular markers show results from explicit randomisation while lines show the evolution according to maximum caliber.}
    \label{fig:WS_var_oscillate_evolution_ensemble_entries}
\end{figure}

The second scenario with varying average number of connections considers a sinusoidal signal $S(t) := K \sin(\omega t)$ which, when positive, indicates connections are added (i.e. some off states of links become on states) and when negative, indicates they are removed (on states become off states). In explicit randomisation by simulations, the amount of added or removed connections at each step $T$ is an integer drawn from a binomial distribution of the same amount of trials as disconnected or connected links respectively and average $|S(T)|$. For maximum caliber applications, the number of added connections, on average over simulations is $\Delta_c(T) = |S(T)|$ if $S(T) > 0$ and $\Delta_c(T) = 0$ otherwise. The number of removed connections is $\Delta_a(T) = |S(T)|$ if $S(T) < 0$ and $\Delta_a(T) = 0$ otherwise. Therefore the constraint on the average change in the number of connections can be written as
\begin{equation}
    \sum_{W_T} \sum_{i,j>i} \left( w_{ij}(T) - w_{ij}(T-1) \right) P(W_T) = \Delta_c(T) - \Delta_a(T) \, ,
    \label{eq:link_variation_sin}
\end{equation}
and the number of changes made in the network configuration becomes
\begin{equation}
    \sum_{W_T} \sum_{i,j>i} | w_{ij}(T) - w_{ij}(T-1) | P(W_T) = 2 + \Delta_c(T) + \Delta_a(T) \, .
    \label{eq:successive_difference_variation_sin}
\end{equation}
Following the same steps as before, the transitions are translated into annihilation and creation probabilities that are the same for every link in the network, with values that can be found by introducing them explicitly into \cref{eq:link_variation_sin,eq:successive_difference_variation_sin}. This results in the system of equations
\begin{equation}
    \begin{aligned}
        \Delta_c(T) - \Delta_a(T) &= (N(N-1)/2 - L) c - a L \\
        2 + \Delta_c(T) + \Delta_a(T) &= (N(N-1)/2 - L) c + a L
    \end{aligned}
\end{equation}
which yields
\begin{equation}
    c = \frac{1 + \Delta_c(T)}{N(N-1)/2 - L} ~~~~~~~~ a = \frac{1 + \Delta_a(T)}{L} \, .
\end{equation}
In \cref{fig:WS_var_oscillate_evolution_ensemble_entries} we show the evolution of selected links starting from ensemble-like initial conditions with a sinusoidal variation of the average number of connections according to simulations and maximum caliber. The values of $K$ are half of the initial average number of connected links of each ensemble-like initial condition and $\omega = 2 \pi / 70$ in all cases. We show the evolution up to a maximum of $500$ steps as the long-term behaviour of all connection probabilities being the same, but oscillating in time, is already reached by then. As in the first case of varying average number of connections, probabilities obtained from maximum caliber and simulations are found to match.

Note that while samples drawn from an ensemble of networks with a constant average number of connections allow for a certain variation in the number of connections of each sample, these are random fluctuations. By changing constraint values, on the other hand, we can control the dynamics of the ensemble average beyond conservation laws, leading to changes in the number of connections of samples due to this dynamic control and fluctuations.

\section{Maximum entropy degree-preserving rewiring}
\label{sec:degree_rewiring}

Consider now a rewiring process such that the degree sequence, that is the number of connections of each node, is conserved. For simplicity of both the explicit randomisation and maximum caliber, as well as the flexibility thereof, we will now consider directed networks with no self-interactions where the link $ij$ is considered different from $ji$. A single step of explicit randomisation consists of selecting a group of two pairs of connected links $ij$ and $kl$ in the network such that all nodes in the group are different and both links $il$ and $kj$ are disconnected. Links $ij$ and $kl$ are then disconnected while $il$ and $kj$ are connected.

From the perspective of maximum caliber, the conservation of the number of connections of each node requires a constraint for each node $j$ defining the change of its in degree,
\begin{equation}
    \sum_{W_T} \sum_{i \neq j} \left( w_{ij}(T) - w_{ij}(T-1) \right) P(W_T) = 0 \, ,
    \label{eq:in_degree_sequence_conservation}
\end{equation}
and one constraint for each node $i$ establishing the change in its out degree
\begin{equation}
    \sum_{W_T} \sum_{j \neq i} \left( w_{ij}(T) - w_{ij}(T-1) \right) P(W_T) = 0 \, .
    \label{eq:out_degree_sequence_conservation}
\end{equation}

The number of changes in the network configuration in a single step of the randomisation is now $4$ instead of $2$ as two connections are removed and two are added. The corresponding constraint is
\begin{equation}
    \sum_{W_T} \sum_{i,j \neq i} | w_{ij}(T) - w_{ij}(T-1) | P(W_T) = 4 \, .
    \label{eq:degree_successive_difference}
\end{equation}

Just as for Watts-Strogatz rewiring, the constraints produce Markov transitions as the constraint functions depend only on the two last network configurations. Introducing Lagrange multipliers $\alpha^{\text{in}}_j$ for \cref{eq:in_degree_sequence_conservation}, $\alpha^{\text{out}}_i$ for \cref{eq:out_degree_sequence_conservation} and $\beta$ for \cref{eq:degree_successive_difference}, we have
\begin{equation}
    \begin{aligned}
        \sum_n \lambda_n & F_n (W_T) \\
        = \sum_{i,j \neq  i} & ( \alpha^{\text{in}}_j + \alpha^{\text{out}}_i ) (w_{ij}(T) - w_{ij}(T-1)) \\ 
          & + \beta |w_{ij}(T) - w_{ij}(T-1)| 
    \end{aligned}
\end{equation}
meaning that \cref{eq:constraint_decomposition} is valid with $\lambda^0_{ij} = \alpha^{\text{out}}_i + \alpha^{\text{in}}_j$, $\lambda^{1}_{ij} = \beta$, $G^{0}_{ij}(w^T_{ij}) = w_{ij}(T) - w_{ij}(T-1)$ and $G^{1}_{ij}(w^T_{ij}) = |w_{ij}(T) - w_{ij}(T-1)|$. Note that the sum over links is carried out as a sum over all non-diagonal elements of the adjacency matrix $i,j\neq i$. By the results from \cref{sec:maximum_caliber_networks} the network transition matrix $M_T$ is a product of independent Markovian link transitions which, following the same steps as in \cref{sec:watts_strogatz}, results in link transition matrices
\begin{equation}
    \begin{aligned}
        m_{ij} :&= P_{ij}(w_{ij}(T) | w_{ij}(T-1)) \\
        &= \begin{pmatrix}
            1 - c_{ij} & a_{ij} \\
            c_{ij} & 1 - a_{ij} 
        \end{pmatrix} \\
        &= \begin{pmatrix}
        \frac{1}{1 + \exp( -\alpha^{\text{out}}_i -\alpha^{\text{in}}_j -\beta))} & \frac{1}{1 + \exp( -\alpha^{\text{out}}_i -\alpha^{\text{in}}_j + \beta)} \\
        \frac{1}{1 + \exp( \alpha^{\text{out}}_i + \alpha^{\text{in}}_j + \beta)} & \frac{1}{1 + \exp( \alpha^{\text{out}}_i + \alpha^{\text{in}}_j - \beta))}            
        \end{pmatrix}
    \end{aligned}
    \label{eq:degree_conserving_transitions}
\end{equation}

Differently from the Watts-Strogatz rewiring process, the transition matrix of each link is different. This makes it extremely difficult to analytically obtain a relation between the annihilation and creation probabilities and the imposed average values. However, as in the binary configuration model, that is the equilibrium case of imposing the degree of each node, \cref{eq:degree_conserving_transitions} gives a functional form of the annihilation and creation probabilities in terms of the Lagrange multipliers which can be adjusted numerically by imposing the constraints of \cref{eq:in_degree_sequence_conservation,eq:out_degree_sequence_conservation,eq:degree_successive_difference} 

\begin{figure}[h!]
    \centering
    \includegraphics[width=\linewidth]{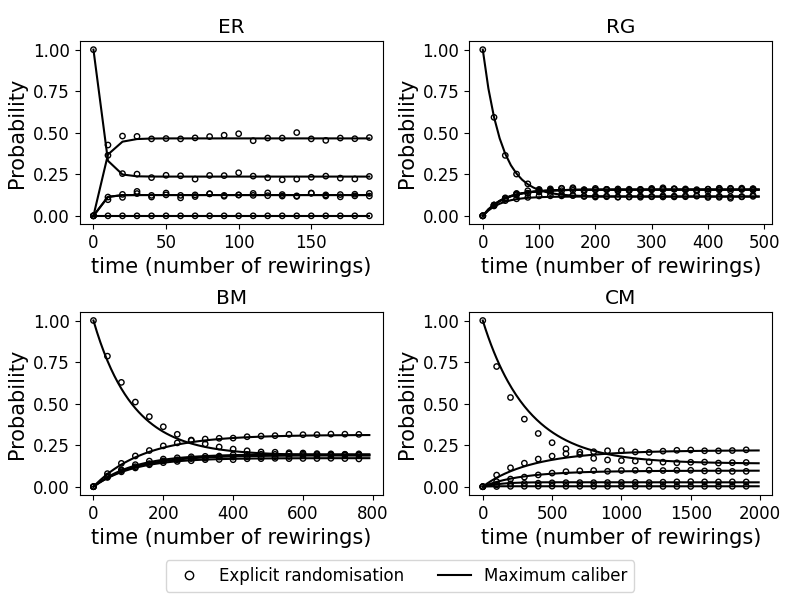}
    \caption{Probability of $5$ different pairs of nodes of being connected as a function of time (number of steps) comparing results from simulations and maximum entropy.}
    \label{fig:degree_rewiring}
\end{figure}

In \cref{fig:degree_rewiring} we show the evolution of connection probabilities over time according to explicit randomisation simulations, in circular markers, and maximum caliber, in full lines, for $5$ pairs of nodes from sample-like initial conditions
The most notable difference between this case and Watts-Strogatz rewiring is that, as the link transition matrices are different for each pair of nodes, trajectories can start from the same connection probability and still evolve differently. There is also a larger difference (due to the number of realisations required) between the simulation and maximum caliber results. We have verified that the stationary connection probability values achieved by individual links (the asymptotic values of \cref{fig:degree_rewiring}) agree with the distribution expected by the directed binary configuration model, $p^{\text{eq}}_{ij} = (1 + \exp(\lambda^{\text{out}}_i + \lambda^{\text{in}}_j))^{-1}$ such that $\sum_i p^{\text{eq}}_{ij} = k^{\text{in}}_j$ is the in degree of node $j$ and $\sum_j p^{\text{eq}}_{ij} = k^{\text{out}}_i$ the out degree of node $i$ obtained from the initial network. This again shows that the distributions resulting from traditional entropy maximisation correspond to equilibrium distributions of a dynamic process defined by the analogous conservation constraints in the context of maximum caliber.

\section{Discussion and conclusions}
\label{sec:discussion_and_conclusions}

In this work, we have applied the principle of maximum caliber of Jaynes to construct the evolution of random network configuration probabilities from constraints representing statistical properties of the evolution. The method is an approach to constructing dynamic processes in the same way that stationary network distributions are obtained by maximising Shannon entropy. The main difference between the method presented here and other applications of maximum caliber is that it obtains individual transition probabilities at different times in the evolution instead of probabilities of entire trajectories. The transition probabilities can then be used to obtain the evolution of a dynamic distribution from desired initial conditions. 

In \cref{sec:maximum_caliber_networks} we show how to obtain such transitions from maximum caliber by replacing the requirement of probability normalisation with a marginalisation, essentially imposing the history of the network evolution. This retains the generality of the method of maximum caliber, in principle allowing for memory-dependent processes. We then highlight specific conditions under which the constraints result in Markov processes and the transitions of the whole network can be described by those of individual links. In \cref{app:maximum_caliber_and_markov} we extend the conditions for Markov processes, and in \cref{app:maximum_entropy_production} we also show that this formulation of maximum caliber can be interpreted as an analogous in information theory to the maximum entropy production principle in the field of thermodynamics, and can be used to strengthen the theoretical basis of the method of entropic dynamics. Next, we focus on particular choices of constraints under which conditions for Markov processes and evolution by individual links apply. In \cref{sec:watts_strogatz} we start with the conservation of the average number of connections in a network, showing that the dynamic distributions that result predict the same connection probabilities as estimated from repetitions of explicit simulations of the rewiring process of Watts and Strogatz. We then modify the constraints in order to represent a controlled variation of the average number of connections, obtaining the same results from simulations of modified Watts-Strogatz randomisation which produce the same average variation. This leads us to conclude that constraints control the evolution of imposed properties rather generally, with conservation resulting as a particular case.  In \cref{sec:degree_rewiring} we apply the same procedure to the conservation of the number of connections of each node in a network, comparing to explicit simulations implementing the same process and once again showing that connection probabilities match.

Our results allow us to conclude that maximum caliber can serve as a useful tool to obtain the evolution of network ensembles undergoing randomisation processes without requiring explicit simulations. It establishes the evolution on the statistical basis of information theory, allowing for the flexibility given by imposing arbitrary constraints that represent properties required of the network evolution. As for future work, three unexplored topics take the spotlight. Firstly, the method as presented here is discrete in time, a limitation that needs to be overcome for the method to be applied to systems continuous in time. Second, in terms of weighted networks, it has been shown that equilibrium maximum entropy networks are better reconstructed by imposing constraints on binary and weighted properties simultaneously. Such constraints in a dynamic context might be applied to the interplay between the network structure of dynamical systems and the dynamics on that structure. Lastly, in terms of memory dependence, the fact that the method naturally incorporates non-Markov processes suggests that it is worthwhile to take a closer look at the perspective provided, especially in the context of complex systems where such effects are paramount. Addressing these challenges would allow for the method to be applied to the study of many real-world complex networks.

\begin{acknowledgments}

We would like to thank Martin Kuffer for discussions and revisions of calculations in this work. Also to Professor Leonid Martyushev and Professor Mario Abadi for their insightful comments and critiques which accompanied us throughout the process of developing the present analysis and writing this article.

\end{acknowledgments}

\appendix

 \section{Connection to maximum entropy production and entropic dynamics}
\label{app:maximum_entropy_production}

In \cref{sec:maximum_caliber_networks} transitions $W_{T-1} \rightarrow W_T$ which essentially update the configuration of the network $W$ at the end of the trajectory were established by maximising the Shannon entropy of the trajectory distribution
\begin{equation}
    S = - \sum_{W_T} P(W_T) \ln(P(W_T))
\end{equation}
subject to constraints
\begin{equation}
    \sum_{W_T} F_n(W_T) P(W_T) = f_n
\end{equation}
of which one is marginalisation $\sum_{W_T} \delta_{W_{T-1}'}(W_{T}) P(W_T) = P(W_{T-1}')$. Although the constraint value $P(W_{T-1}')$ is the trajectory distribution at the previous number of steps, it is also arbitrary and for all purposes of the maximisation, fixed. One can then also maximise the entropy production
\begin{equation}
    \begin{aligned}
        \Delta S_T =& - \sum_{W_T} P(W_T) \ln(P(W_T)) \\
                    & + \sum_{W_{T-1}} P(W_{T-1}) \ln(P(W_{T-1})) \\
                   =& - \sum_{W, W_{T-1}} P(W_{T-1}, W) \ln(P(W(T), W_{T-1})) \\
                    & + \sum_{W, W_{T-1}} P(W_{T-1}, W) \ln(P(W_{T-1})) \\
                   =& - \sum_{W, W_{T-1}} P(W | W_{T-1}) P(W_{T-1}) \ln(P(W | W_{T-1})) \, .
    \end{aligned}
    \label{eq:entropy_production}
\end{equation}

As $P(W | W_{T-1})$ holds $P(W_{T-1})$ fixed, we can maximise the entropy production with respect to the transition probability instead of the trajectory distribution, writing the constraints as
\begin{equation}
    \begin{aligned}
        \sum_{W, W_{T-1}} P(W | W_{T-1}) F_n(W, W_{T-1}) P(W_{T-1}) &= f_n \\
        \sum_{W, W_{T-1}} \delta_{W_{T-1}'}(W_{T}) P(W | W_{T-1}) = \sum_W P(W | W_{T-1}') &=  1 \, .
    \end{aligned}
\end{equation}
This defines a Lagrangian
\begin{equation}
    \begin{aligned}
        & \mathcal{L} = - \sum_{W, W_{T-1}} P(W|W_{T-1}) P(W_{T-1}) \ln( P(W | W_{T-1}) ) \\
                    &+ \sum_{W_{T-1}} \lambda_{W_{T-1}} \left( 1 - \sum_{W} P(W | W_{T-1}) \right) \\
                    &+ \sum_n \lambda_n \left( f_n - \sum_{\substack{W \\ W_{T-1}}} P(W | W_{T-1}) F_n(W, W_{T-1}) P(W_{T-1}) \right)
    \end{aligned}
\end{equation}
maximised by finding the roots of
\begin{equation}
    \begin{aligned}
        & \frac{\partial \mathcal{L}}{\partial P(W|W_{T-1})} = - \lambda_{W_{T-1}}  \\ 
        & -  P(W_{T-1}) \left[ \ln(P(W|W_{T-1}) + 1 + \sum_n \lambda_n F_n(W,W_{T-1}) \right] 
    \end{aligned}
\end{equation}
in terms of $P(W | W_{T-1})$. This yields
\begin{equation}
    \begin{aligned}
        & P(W|W_{T-1}) = \\
        & \exp \left(-\left[ \lambda_{W_{T-1}} / P(W_{T-1}) + 1 + \sum_n \lambda_n F_n(W,W_{T-1}) \right] \right) \, .
    \end{aligned}
\end{equation}
Imposing marginalisation $\sum_{W} P(W|W_{T-1}) = 1$,
\begin{equation}
    \begin{aligned}
        & \exp \left( \lambda_{W_{T-1}} / P(W_{T-1}) + 1 \right) = \\
        & ~~~~~~~~ \sum_{W} \exp \left( -\sum_n \lambda_n F_n(W, W_{T-1}) \right) \\
        \Rightarrow & P(W | W_{T-1}) = \frac{ \exp \left( -\sum_n \lambda_n F_n(W, W_{T-1}) \right) }{ \sum_{W'} \exp \left( -\sum_n \lambda_n F_n(W', W_{T-1}) \right) }
    \end{aligned}
    \label{eq:entropy_production_transitions}
\end{equation}

As \cref{eq:entropy_production_transitions} matches the result obtained in \cref{eq:maxcal_transitions}, maximum caliber in combination with marginalisation constraints is equivalent to maximisation of the entropy production defined in \cref{eq:entropy_production}. Moreover, the same functional form is reached if the "entropy production" is arbitrarily defined as
\begin{equation}
    -\sum_{W,W_{T-1}} P(W | W_{T-1}) \ln(P(W|W_{T-1}))
    \label{eq:entropic_dynamics_entropy}
\end{equation}
with constraints
\begin{equation}
    \sum_{W, W_{T-1}} P(W | W_{T-1}) F_n(W, W_{T-1}) = g_n
    \label{eq:entropic_dynamics_constraints}
\end{equation}
independently of the constraint values $g_n$. We also know that there must exist constraint values $g_n$ such that the resulting transition probabilities, and not just the functional form, are the same as derived from maximum caliber. These can be constructed by combining the functional form of the transitions, which is independent of the method used, with the Lagrange multipliers resulting from maximum caliber. This fully defines the transition probabilities on the left-hand side of \cref{eq:entropic_dynamics_constraints}, which allows the construction of the constraint values on the right-hand side.   
 
Note that for the case of Markov processes, we can write $P(W(T) | W_{T-1}) = P(W(T) | W(T-1))$. In particular \cref{eq:entropic_dynamics_entropy,eq:entropic_dynamics_constraints} under this condition result in the formulation used in entropic dynamics. This shows that the latter method yields the correct results, and sets it on the more solid foundations of Maximum caliber.

\section{Maximum caliber and Markov processes}
\label{app:markov_processes}
\label{app:maximum_caliber_and_markov}

We have established in \cref{sec:maximum_caliber_networks} that when constraint functions depend only on two successive states $W(T-1)$ and $W(T)$ of the network, the resulting transitions $M_T$ define a Markov process. However, this condition can be somewhat loosened by considering linear combinations of constraints. For example, consider a more typical constraint of maximum caliber defining the average number of connections $C(T)$ over a trajectory
\begin{equation}
    C(T) = \sum_{W_T} \sum_{ij} \sum_{0 \leq t \leq T} w_{ij}(t) P(W_T) \, .
    \label{eq:trajectory_total}
\end{equation}
For trajectories one step shorter, the same constraint is
\begin{equation}
    \begin{aligned}
    C(T-1)  &= \sum_{W_{T-1}} \sum_{ij} \sum_{0 \leq t \leq T-1} w_{ij}(t) P(W_{T-1}) \\
            &= \sum_{W_T} \sum_{ij} \sum_{0 \leq t \leq T-1} w_{ij}(t) P(W_T) \, .
    \end{aligned}
\end{equation}
which can be subtracted from \cref{eq:trajectory_total} to yield the average number of connections $L(T)$ at time $T$
\begin{equation}
    C(T) - C(T-1) = \sum_{W_T} \sum_{ij} w_{ij}(T) P(W_T) = L(T) \, .
    \label{eq:subtracted_number_of_links}
\end{equation}
Considering \cref{eq:subtracted_number_of_links} for $T-1$ and $T$, these can also be subtracted, obtaining the constraints used for \cref{sec:watts_strogatz}
\begin{equation}
    L(T) - L(T-1) = \sum_{W_T} \sum_{ij} ( w_{ij}(T) - w_{ij}(T-1) ) P(W_T) \, .
    \label{eq:second_subtracted_number_of_links}
\end{equation}
On the other hand, consider the case where constraints over trajectories include coefficients that depend on the length of the trajectory, for example
\begin{equation}
     \sum_{W_T} \sum_{ij} \left( \sum_{t=0}^T A_T e^{-t} w_{ij}(t) \right) P(W_T) = C(T) \, .
\end{equation}
When we attempt to construct the instantaneous constraint by difference of two successive times, we find that the result again depends on the whole trajectory,

\begin{equation}
    \begin{aligned}
        C(T) - C(T-1) & = \\ 
        \sum_{W_T} \sum_{ij} & \left[ A_T e^{-T} w_{ij}(T) + \right. \\ 
        & ~ +\sum_{t=0}^{T-1} \left. (A_T - A_{T-1}) e^{-t} w_{ij}(t) \right] P(W_T)\, ,
    \end{aligned}
\end{equation}
and the same is true for higher-order differences, suggesting that the resulting process is not Markov. 

\bibliography{ref}

\end{document}